\documentclass[12pt,oneside]{article}

\usepackage[left=3.4cm,top=3.6cm,bottom=3.6cm,right=3.4cm,head=0cm,foot=0.7cm]{geometry}

\usepackage{amsmath,amssymb}

\newcommand{\field}[1]{\mathbb{#1}}

\date{}

\title{Reconceptualising equilibrium in Boltzmannian statistical mechanics and characterising its existence}
\author{Professor Charlotte Werndl\\
 (University of Salzburg and London School of Economics)\\and Professor Roman Frigg \\(London School of Economics)}
 \date{This article is forthcoming in:
Studies in History and Philosophy of Modern Physics
http://www.journals.elsevier.com/studies-in-history-andphilosophy-of-science-part-b-studies-in-history-and-philosophyof-modern-physics/}

\begin{document}

\newtheorem{theorem}{Theorem}

\maketitle

\author

\begin{abstract}
In Boltzmannian statistical mechanics macro-states supervene on micro-states. This leads to a partitioning of the state space of a system into regions of macroscopically indistinguishable micro-states. The largest of these regions is singled out as the equilibrium region of the system. What justifies this association? We review currently available answers to this question and find them wanting both for conceptual and for technical reasons. We propose a new conception of equilibrium and prove a mathematical theorem which establishes in full generality -- i.e. without making any assumptions about the system's dynamics or the nature of the interactions between its components -- that the equilibrium macro-region is the largest macro-region. We then turn to the question of the approach to equilibrium, of which there exists no satisfactory general answer so far. In our account, this question is replaced by the question when an equilibrium state exists. We prove another -- again fully general -- theorem providing necessary and sufficient conditions for the existence of an equilibrium state. This theorem changes the way in which the question of the approach to equilibrium should be discussed: rather than launching a search for a crucial factor (such as ergodicity or typicality), the focus should be on finding triplets of macro-variables, dynamical conditions, and effective state spaces that satisfy the conditions of the theorem.
\end{abstract}

\tableofcontents

\newpage

\section{Introduction}
The core posit of Boltzmannian statistical mechanics (BSM) is that macro-states supervene on micro-states. This leads to a partitioning of the state space of a system into regions of macroscopically indistinguishable micro-states, where by `macroscopically indistinguishable' we mean indistinguishable with respect to macroscropic properties such as thermodynamic properties. These regions are called \textit{macro-regions}. The largest of these macro-regions is commonly singled out as the system's equilibrium region. What justifies the association of equilibrium with the macro-state corresponding to the largest macro-region?\\

After briefly introducing the main elements of BSM (Section 2) and illustrating them with three examples, we scrutinise common answers that have been given to this question. We find these wanting both for conceptual and for technical reasons (Section 3). This prompts the search for an alternative answer. This answer cannot be found by revising any of the received approaches, and so we propose a new definition of equilibrium. While previous approaches sought to define equilibrium in terms of micro-mechanical properties, our definition is modelled on the thermodynamic conception of equilibrium, and also incorporates what has become known as the `minus first law' of thermodynamics (TD) (Section 4).\\

The new conception of equilibrium is not only free from the conceptual and technical difficulties of earlier notions; it also provides the spring-board for a general answer to our initial problem. We prove a mathematical theorem which establishes in full generality that the equilibrium macro-region is the largest macro-region (in a requisite sense). The proof is mathematically rigorous and the theorem is completely general in that it makes \textit{no} assumptions either about the system's dynamics or the nature of the interactions between the system's components (Section 5). \\

We then turn to the question of the approach to equilibrium, to which there exists no satisfactory general answer. In our account, this question is replaced by the question: under what circumstances does an equilibrium state exists? We point out that for an equilibrium to exist three factors need to cooperate: the choice of macro-variables, the dynamics of the system, and the choice of the effective state space. We then prove a theorem providing fully general necessary and sufficient conditions for the existence of an equilibrium state. This theorem changes the way in which the problem of the approach to equilibrium should be discussed: rather than launching a search for one crucial factor (such as ergodicity or typicality), the focus should be on finding triplets of macro-variables, dynamical conditions, and effective state spaces that satisfy the conditions of the theorem. This gives the discussion of equilibrium a new direction (Section 6).

\section{Boltzmannian Statistical Mechanics}\label{BSM}

We begin with a brief summary of the apparatus of BSM. This is mainly to introduce notation and state a few crucial results; for detailed introductions to BSM we refer the reader to Frigg (2008) and Uffink (2007). We then introduce three examples that will guide us through our discussion and serve as illustrations of the general claims we make in later sections. The reliance on three different examples is not owed to a preference for abundance. Discussions of BSM have all too often been distorted, and indeed misled, by an all too narrow focus on the dilute gas. Contrasting the dilute gas (our first example) with the baker's gas and the Kac-ring (our second and third examples) widens the focus and helps illustrate the general claims we make in later sections.

\subsection{The Framework of Boltzmannian Statistical Mechanics}
A  system in statistical mechanics has the mathematical structure of a \emph{measure-preserving deterministic dynamical system} $(X,\Sigma_{X},\mu_{X}, T_{t})$. $X$ is the set representing all possible \emph{micro-states}; $\Sigma_{X}$ is a $\sigma$-algebra of  subsets of $X$; the \emph{evolution function} $T_{t}:X\rightarrow X$, $t\in\field{R}$ (continuous time) or $\field{Z}$ (discrete time), is a measurable function in $(t,x)$ such that $T_{t_{1}+t_{2}}(x)=T_{t_{2}}(T_{t_{1}}(x))$ for all $x\in X$ and all $t_{1},t_{2}\in\field{R}$ or $\field{Z}$; $\mu_{X}$ is a measure on $\Sigma_{X}$ that it is invariant under the dynamics: $\mu_{X}(T_{t}(A))=\mu_{X}(A)$ for all $A\in\Sigma_{X}$ and all $t$.\footnote{At this point the measure of $X$ is allowed to be infinite (hence there is no requirement that the measure is normalized).} The \textit{solution} through $x$, $x\in X$, is the function $s_{x}:\field{R}\rightarrow X$ or $s_{x}:\field{Z}\rightarrow X$, $s_{x}(t)=T_{t}(x)$.\\

At the macro level the system is characterised by a set of \emph{macro-variables} $\{v_{1}, ..., v_{l}\}$ ($l \in \field{N}$). These variables are measurable functions $v_{i}:X\rightarrow \field{V}_{i}$, associating a value with each point in $X$. We use capital letters $V_{i}$ to denote the values of $v_{i}$. A particular set of values $\{V_{1}, ..., V_{l}\}$ defines a \emph{macro-state} $M_{V_{1}, \ldots, V_{l}}$.
We only write `$M$' rather than `$M_{V_{1}, ..., V_{l}}$' if the specific values $V_{i}$ do not matter. For now all we need is the general definition of macro-variables. We will discuss them in more detail in Section \ref{HT}, where we will see that the choice of a set of macro-variables is a subtle matter of considerable importance and that the nature and even existence of an equilibrium state crucially depends on it. \\

The central philosophical posit of BSM is \emph{supervenience}: macro-states supervene on micro-states. This implies that a system's micro-state uniquely determines its macro-state. This determination relation will be many-to-one.
For this reason every macro-state $M$ is associated with a macro-region consisting of all micro-states for which the system is in $M$. An important yet often neglected issue is on what space macro-regions are defined. The obvious option would be $X$, but often this is not what happens. In fact, in many cases macro-regions are defined on a subspace $Z \subseteq X$. Intuitively speaking, $Z$ is a subset whose states evolve into the same equilibrium macro-state. In the case of a dilute gas with $N$ particles, for instance, $X$ is the $6N$-dimensional space of all position and momenta, while $Z$ is the $6N-1$ dimensional energy hypersurface. We call $X$ the \textit{full state space} and $Z$ the \textit{effective state space} of the system. The \emph{macro-region $Z_{M}$} corresponding to \emph{macro-state $M$ relative to $Z$} can then be defined as the set of all $x \in Z$ for which  $M$ supervenes on $x$. A set of macro-states relative to $Z$ is complete iff (if and only if) it contains all states of $Z$. The members of a complete set of macro-regions $Z_{M}$ form a partition of $Z$ (i.e. the $Z_{M}$  do not overlap and jointly cover $Z$).\\

The correct choice of $Z$ depends on the system under investigation, and has to be determined on a case-by-case basis. We return to this point in Subsection~\ref{HT}. There is one general constraint on such a choice, though, that needs to be mentioned now.
Since a system can never leave the partition of macro-regions,
$Z$ must be mapped onto itself under $T_{t}$. Then the sigma algebra can be restricted to $Z$ and one considers a measure on $Z$ which is invariant under the dynamics and where the measure is normalized, i.e. $\mu_{Z}(Z)=1$.\footnote{The dynamics is given by the evolution equations restricted to $Z$, and we follow the dynamical systems literature in denoting it again by $T_{t}$.} In this way one obtains the measure-preserving dynamical system $(Z,\Sigma_{Z},\mu_{Z},T_{t})$  with a normalized measure $\mu_{Z}$. $(Z,\Sigma_{Z},\mu_{Z},T_{t})$ is called the \emph{effective system} (as opposed to the \emph{full system} $(X,\Sigma_{X},\mu_{Z},T_{t})$).
\\

The \emph{Boltzmann entropy} of a \emph{macro-state} $M$ relative to $Z$ is $S_{B}(M)\,:=\,k_{B}\log[\mu_{Z}(Z_{M})]$ ($k_{B}$ is the Boltzmann constant). The Boltzmann entropy of a \emph{system} at time $t$, $S_{B}(t)$, is the entropy of the macro-state the system is in at $t$ relative to $Z$: $S_{_{B}}(t):=S_{_{B}}(M_{x(t)})$, where $x(t)$ is the system's micro-state at $t$ and $M_{x(t)}$ is the macro-state supervening on $x(t)$.\\

One of the macro-regions is singled out as corresponding to the \emph{equilibrium state} of the system relative to $Z$. A crucial aspect of the standard presentation of BSM is that equilibrium corresponds to the largest macro-region (measured in terms of $\mu_{Z}$). In fact, this is often used as a criterion to define equilibrium: the equilibrium state relative to $Z$ is simply the one that is associated with the largest macro-region. Since the logarithm is a monotonic function, the equilibrium state is also the one with the largest Boltzmann entropy.\\

\subsection{Example 1: The Dilute Gas}
Consider a system consisting of $N$ particles in a finite container isolated from the environment. The  \emph{micro-state} of the system is specified by a point $x=(q,p)$ in the $6N$-dimensional set of possible position and momentum coordinates $\Gamma$. So $\Gamma$ is the $X$ of the gas. The dynamics of the system is determined by its classical Hamiltonian $H(x)$. Energy is preserved and therefore the motion is confined to the $6N-1$ dimensional energy hypersurface $\Gamma_{E}$ defined by $H(x)=E$, where $E$ is the energy value. So $\Gamma_{E}$ is the $Z$ of the gas. The solutions of the equations of motion are given by the flow $T_{t}$ on $\Gamma_{E}$, where $T_{t}(x)$ is the state into which $x \in \Gamma_{E}$ evolves after time $t$ has elapsed. $\Sigma_{E}$ is the standard Lebesgue-$\sigma$-algebra. $\Gamma$ is endowed with the Lebesgue measure $\lambda$, which is preserved under $T_{t}$.  A measure $\mu_{E}$ on $\Gamma_{E}$ can be defined which is preserved as well and is normalised, i.e.\  $\mu_{E}(\Gamma_{E})=1$ (cf.\ Frigg 2008, 104). $(\Gamma_{E},\Sigma_{E},\mu_{E},T_{t})$ is the effective measure-preserving dynamical system of the gas.\\

The macro-states usually considered arise as follows: the state of one particle is determined by a point in its $6$-dimensional state space $\gamma$, and the state of system of $N$ identical particles is determined by $N$ points in this space. Since the system is confined to a finite container and has constant energy $E$, only a finite part of $\gamma$ is accessible. One then partitions the accessible part of $\gamma $ into cells of equal size $\delta\omega$ whose dividing lines run parallel to the position and momentum axes. The result is a finite partition $\Omega_{dg} := \{\omega^{dg}_{1}, \, ..., \omega^{dg}_{l}\}$, $l\in\field{N}$ (where the subscript `\emph{dg}' stands for `dilute gas'). The cell in which a particle's state lies is its \textit{coarse-grained micro-state}. The coarse-grained micro-state of the entire gas, called an \emph{arrangement}, is given by a specification of the coarse-grained micro-state of each of particle.\\

A specification of the `occupation number' of each cell is know as a \textit{distribution} $D_{dg}=(N_{1}, N_{2},\ldots,N_{l})$, where $N_{i}$ is the number of particles whose state is in cell $\omega^{dg}_{i}$. Since macro-properties are fixed by the distribution, the \emph{macro-states correspond to the different distributions}. Each distribution is compatible with several arrangements, and the number $G(D_{dg})$ of arrangements compatible with a given distribution $D$ is $G(D_{dg}) = N!\, / \, N_1!N_2!\ldots,N_l!$.\\

Every micro-state $x$ of $\Gamma_{E}$ is associated with exactly one distribution $D_{dg}(x)$. One then defines the \emph{macro-region} $\Gamma_{D_{dg}}$ as the set of all $x$ that are associated with macro-state $D_{dg}$: $\Gamma_{D_{dg}} = \{x\in\Gamma_E\,\,:\,\,D_{dg}(x)=D_{dg}\}$. The \emph{equilibrium macro-region} is defined as the macro-region of largest measure $\mu_{E}$.\\

In his famous 1977 paper Boltzmann provided an argument to determine the equilibrium distribution, nowadays referred to as the \emph{combinatorial argument}. He assumed that the energy $e_i$ of particle $i$ depends only on the cell in which it is located and that thus
$E=\sum_{i=1}^{l}N_{i}e_{i}$. Assuming that the number of cells in $\Omega_{dg}$ is small compared to the number of particles, Boltzmann showed that $\mu_{E}(\Gamma_{D_{dg}})$ is maximal when
\begin{equation}\label{rig}
N_i = \gamma e^{\lambda e_{i}},
\end{equation}
where $\gamma$ and $\lambda$ are parameters which depend on $N$ and $E$. This is the \emph{discrete version of the Maxwell-Boltzmann distribution}. Thus the equilibrium macro-state corresponds to the Maxwell-Boltzmann distribution.\\

What $(\ref{rig})$ gives us is the distribution of largest size (for the Lebesgue measure) on the $6N$-dimensional shell-like domain $\Gamma_{ES}$ specified by the condition that $E=\sum_{i=1}^{l}N_{i}e_{i}$. It does \emph{not} give us the macro-region of maximal size (i.e., the distribution with the largest measure $\mu_{E}$ on the $6N-1$ dimensional $\Gamma_{E}$). It is then typically assumed that the possible distributions and the proportion of the different distributions would not change if macro-states were instead defined on $\Gamma_{E}$, which yields the result that the equilibrium region is the largest region on $\Gamma_{E}$. As Ehrenfest and Ehrenfest (1959, 30) stress, this assumption is in need of further justification. We grant, for the sake of argument, that such a justification can be given and that the equilibrium macro-region constructed in the above manner is the largest region of $\Gamma_{E}$.

\subsection{Example 2: The Baker's Gas}

The baker's gas consists of $N$ identical particles that evolve (independently of each other) according to the baker's transformation (cf.\ Lavis 2005). Its \emph{micro-states} are of the form $b=(b_{1},c_{1},\ldots,b_{N},c_{N})$, where $b_{i} \in [0,1]$ is the momentum and $c_{i}\in[0,1]$ is the position coordinate of the $i$-th particle. The possible micro-states are the set $B=[0,1]^{2N}$, which is both the $X$ and $Z$ of the baker's gas. Time is discrete and the evolution after one time step is given by applying to each coordinate the baker's transformation. That is, the state $b=(\ldots b_{i},c_{i}\ldots)$ evolves into the state $\Lambda(b)=(\ldots \theta(b_{i},c_{i})\ldots)$, where
\begin{equation}
\theta(b_{i},c_i)=2b_i,\frac{c_i}{2}\,\,\textnormal{if}\,\,0\leq b_i\leq \frac{1}{2}\,\,\textnormal{and}\,\,2b_{i}-1,\frac{c_{i}+1}{2}\,\,\textnormal{otherwise}.
\end{equation}
$\Sigma_{B}$ is the Lebesgue-$\sigma$-algebra of $B$ and, intuitively speaking, consists of all subsets of $[0,1]^{2N}$. $B$ is endowed with the Lebesgue measure $\mu_{B}$, which is preserved under $\Lambda$. $(B, \Sigma_{B}, \Lambda_{t},\mu_{B})$, where $\Lambda_{t}$ is the $t$-th iterate of $\Lambda$, is a measure-preserving deterministic system describing the behaviour of the baker's gas.\\

The macro-states usually considered arise by applying the same recipe as in Example 1. One starts by partitioning the unit square (the state space for one particle) into cells of equal size $\delta\omega$ whose dividing lines run parallel to the position and momentum axes. This results into a finite partition $\Omega_{bg} := \{\omega_{1}^{bg}, \, ..., \omega_{l}^{bg}\}$, $l\in\field{N}$ (where `\emph{bg}' stands for `baker's gas'). The coarse-grained micro-state of a particle is the cell in which a particle's state lies. An \emph{arrangement} is given by a specification of the coarse-grained micro-state of all the particles.\\

As above, the macro-properties of a system depend only on how many particles there are in each cell and not on which particles these are. That is, they only depend on the distribution $D_{bg}=(N_{1}, N_{2},\ldots,N_{l})$, where $N_{i}$ is the number of particles in cell $\omega_{i}$, and the \emph{distributions $D_{bg}$ are the macro-states}. The number $G(D_{bg})$ of arrangements that lead to the same distribution $D_{bg}$ is $G(D_{bg}) = N!\, / \, N_1!N_2!\ldots,N_l!$. A macro-region $B_{D_{bg}}$ is defined as the set of micro-states that lead to the distribution $D_{bg}$.\\

In keeping with the basic posit of BSM, the \emph{equilibrium} macro-state is defined as the macro-state corresponding to the largest macro-region (measured with $\mu_{B}$). Hence the equilibrium state is characterised by the \emph{uniform distribution} where $N_i=N/l$ for all $i$.\footnote{To make sure that this equilibrium macro-state is unique, we assume that $N$ is a multiple of $l$.}

\subsection{Example 3: The Kac-Ring}

The Kac-ring model consists of an even number $N$ of sites distributed equidistantly around a circle. On each site there is a spin, which can be in states up ($u$) or down ($d$). Hence the one spin state space is $\{u, \, d \}$.  A \emph{micro-state} $k$ of the ring is a specific combination of up and down spin for all sites, and the full state space $K$ consist of all combinations of up and down spins (i.e., of $2^{N}$ elements). $K$ is both the $X$ and $Z$ of the Kac-ring. There are $s$, $1\leq s\leq N-1$, spin flippers distributed at some of the midpoints between the spins. The dynamics $\kappa$ rotates the spins one spin-site in the clockwise direction, and when the spins pass through a spin flipper, they change their direction. The probability measure usually considered is the uniform measure $\mu_{K}$ on $K$. $(K, P(K), \kappa^{t},\mu_{K})$, where $\kappa^{t}$ is the $t$-th iterate of $\kappa$ and $P(K)$ is the power set of $K$, is a measure-preserving deterministic system describing the behaviour of the spins (cf.\ Bricmont 2001; Lavis 2008).\\

The \emph{macro-states} usually considered are the \emph{total number of up spins} and will be labelled as $M^{K}_{i}$, where $i$ denotes the total number of up spins, $0\leq i\leq N$. As above, the \emph{macro-regions} $K_{i}$ are defined as the set of micro-states leading to the macro-state $M^{K}_{i}$. There are $N!/ i!(N-i)!$ micro-states which give rise to the same macro-state $M^{K}_{i}$. It can be shown that the \emph{equilibrium macro-state}, i.e.\ the macro-state whose macro-region is of largest size, is $M^{K}_{N/2}$, the state in which half of the spins are up and half down.\\

We can again describe the equilibrium state for the Kac-ring in terms of a distribution. Since the one spin state space $\{u, \, d \}$ is discrete (in contrast to the dilute gas and the baker's gas where it is continuous), we can regard the states $u$ and $d$ as corresponding to the cells $\omega_{i}$ of the gas. Then the equilibrium distribution is the even distribution (the same number of spins are in $u$ and $d$). So, as in the case of the baker's gas, the equilibrium distribution is \textit{not} the Maxwell-Boltzmann distribution but the even distribution.

\section{Scrutinising the Standard Conception of Equilibrium}

As we have seen in the last section, the Boltzmannian approach associates equilibrium with the largest macro-region. This is taken to be constitutive of equilibrium: the equilibrium state by definition is the state that is associated with the largest macro-region. We call this the \textit{standard conception of equilibrium}. This conception raises two fundamental questions:

\begin{quote}

\textit{Question 1: Justification}. Why is the equilibrium state defined as the state with the largest macro-region? There is no obvious connection between equilibrium and having a large state space measure and therefore this association needs to be justified.

\textit{Question 2: Approach to Equilibrium}. Under what conditions do systems approach equilibrium? One expects systems to approach equilibrium, but not all systems do. This raises the question what dynamical conditions a system has to satisfy for the approach to equilibrium to take place.

\end{quote}

\noindent In this section we discuss currently available answers to both questions and reach the conclusion that none fits the bill. Question 1 is addressed in Subsections \ref{Disambiguation}, \ref{D-Conceptual} and \ref{D-Formal}; Question 2 is discussed in  Subsection \ref{Approach}.

\subsection{Disambiguation: Dominance and Prevalence}\label{Disambiguation}
As we have seen above, the standard version of BSM associates equilibrium with the largest macro-region. But the notion of the `largest macro-region' is ambiguous. It allows for two different readings, which are, however, rarely distinguished clearly. We call these two readings \textit{dominance} and \textit{prevalence} respectively.\\

The first reading is often assumed in the philosophical literature on BSM and takes `largest' to mean that the equilibrium macro-region fills almost the entire effective state space, which, in that context, is typically taken to be the energy hypersurface. This notion of equilibrium can be found, for instance, in Goldstein when he insist that the energy hypersurface `consists almost entirely of phase points in the equilibrium macro-state, with ridiculously few exceptions' (Goldstein 2001, 43).\footnote{The same conception of equilibrium can also be found in: Albert (2000), Bricmont (2001), Goldstein and Lebowitz (2004), Lebowitz (1993a, 1993b) and Penrose (1989).} For the discussion to follow it is helpful to have a formal and more general rendering of this idea. We say that $Z_{M_{eq}}$ is \emph{$\beta$-dominant}
iff $\mu_{Z}(Z_{M_{eq}}) \geq\beta$ for \textit{a particular} $\beta\in (\frac{1}{2},1]$. So if we say that a given $Z_{M_{eq}}$ is $\beta$-dominant, this presupposes that a particular value for $\beta$ is specified (for instance that the $Z_{M_{eq}}$ is $3/4$-dominant). This implies that if a $Z_{M_{eq}}$ is $\beta'$-dominant, then it is in fact also $\beta$-dominant for all $\beta$ in $(1/2, \, \beta')$. If, for instance, $Z_{M_{eq}}$ is $3/4$-dominant, then it is also $2/3$-dominant. Often we are interested in the largest $\beta$ for which $Z_{M_{eq}}$ is $\beta$-dominant. If the largest $\beta$ is close to one, then we retrieve Goldstein's definition of equilibrium.\\

The second reading is often appealed to in calculations and takes `largest' to mean `larger than any other macro-region' (cf. Boltzmann 1877; Bricmont 2001). That is, equilibrium is defined by the condition: $\mu_{Z}(Z_{M_{eq}})>\mu_{Z}(Z_M)$ for any macro-region $M$ with $M \neq M_{eq}$. Again, a formal version of this condition will be useful later on. We say that $Z_{M_{eq}}$ is \emph{$\delta$-prevalent} iff
$\min_{M \neq M_{eq}} [\mu_{Z}(Z_{M_{eq}}) -\mu_{Z}(Z_{M})]\geq\delta$ for a real number $\delta > 0$.
 As above, this presupposes that a particular value for $\delta$ is chosen, and if a $Z_{M_{eq}}$ is $\delta'$-prevalent, then it is also $\delta$-prevalent for all  $\delta$ in $(0, \, \delta')$.\\

At this point we do not aim to adjudicate between these different definitions. We would like to point out, however, that they are not equivalent: $\beta$-dominance implies $\delta$-prevalence, but the converse fails. More specifically: for all $\beta$, if $Z_{M_{eq}}$ is $\beta$-dominant, then it is also $\delta$-prevalent for all $\delta$ in $(0, 2\beta-1]$. In other words, whenever an equilibrium macro-region is $\beta$-dominant, there exists a range of values for $\delta$ so that the macro-region is also $\delta$-prevalent for these values. This is intuitively clear because a macro-region that takes up more than half of $Z$ is also larger than any other macro-region. By contrast, a macro-region that is larger than any other macro-region need not take up more than half of $Z$. So if $Z_{M_{eq}}$ is $\delta$-prevalent, there need not be a $\beta$ so that $Z_{M_{eq}}$ is also $\beta$-dominant. In fact, if there is a large number of macro-regions, the largest macro-region can be relatively small compared to $Z$.\\

This point is often overlooked. As we have seen, many accounts of SM are committed to the view that the equilibrium macro-region is $\beta$-dominant (for a value of $\beta$ close to one). However, calculations usually only establish $\delta$-prevalence. This problem is then often `resolved' by simply brushing the difference between the two notions under the rug and assuming that a $\delta$-prevalent state is also $\beta$-dominant. For instance, Penrose (1989, 403) and Goldstein (2001, 43) support the claim that the equilibrium state fills up almost the entire state space by calculating that the ratio between the measure of the equilibrium macro-region and the macro-region of a standard non-equilibrium state is of order  $10^{N}$. But this amounts to inferring $\beta$-dominance from $\delta$-prevalence.

\subsection{Defining Equilibrium: Conceptual Quandaries}\label{D-Conceptual}

The notion that equilibrium is defined by the largest macro-region (where `largest' can mean either $\beta$-dominant or $\delta$-prevalent) is deeply entrenched in BSM, and it is shared by rivalling versions of BSM. Those who favour an account of BSM based on ergodic theory have to assume that $Z_{M_{eq}}$ is large because otherwise the system would not spend most of its time in equilibrium (see, for instance, Frigg and Werndl 2011a, 2012a, 2012b). Those who see the approach to equilibrium as result of some sort of probabilistic dynamics assume that $Z_{M_{eq}}$ is large because they assign probabilities to macro-states that are proportional to $\mu_{E}(\Gamma_{M})$ and equilibrium comes out as the most likely state only if the equilibrium macro-region is large (e.g.\ Boltzmann 1877). Proponents of the typicality approach see $\beta$-dominance (for $\beta$ close to $1$) as the key ingredient in explaining the approach to equilibrium and sometimes even seem to argue that systems approach equilibrium \textit{because} the equilibrium region takes up nearly all of state space (e.g.\ Goldstein 2001; Goldstein and Lebowitz 2004).\\

However, the connection between equilibrium and large macro-regions is not conceptual: there is nothing in the \textit{concept} of equilibrium tying it to either $\beta$-dominance or $\delta$-prevalence. Hence, irrespective of their merits, all these accounts have to answer the same fundamental question: what justifies the association of equilibrium with the largest macro-region?\\

A prominent answer originates in Boltzmann's 1877 paper: \emph{equilibrium corresponds to the macro-state that is compatible with the largest
number of micro-states}. Boltzmann then shows for dilute gases that equilibrium thus defined is $\delta$-prevalent and that it is characterised by the Maxwell-Boltzmann distribution (see Example 1 above). This way of thinking about equilibrium also seems to be at work in the baker's gas (Example 2) and the Kac-ring (Example 3), where equilibrium is associated with the largest macro-region.\\

This justificatory strategy faces a serious problem: the absence of a conceptual connection with the thermodynamic (TD) notion of equilibrium. The following is a typical TD textbook definition of equilibrium: `A thermodynamic system is in equilibrium when none of its thermodynamic properties are changing with time [...]' (Reiss 1996, 3). An isolated system converges to this state when left to its own and it never leaves it once it has been reached (Callender 2001; Uffink 2001). The problem is that there is simply no \emph{conceptual} connection between this notion of equilibrium and the idea that the equilibrium macro-state is the one that is compatible with the largest number of micro-states. This is a problem for anyone who sees BSM as a reductionist enterprise.\footnote{And while the precise contours of the reduction of TD to SM remain controversial, we are not aware of any contributors who maintain radical anti-reductionism.} \\

One might reply that since $Z_{M_{eq}}$ is the largest subset of  $Z$, systems approach equilibrium and spend most of their time in $Z_{M_{eq}}$. This shows that the BSM definition of equilibrium is a good approximation to the TD definition. This is \emph{not} true in general (Frigg 2010a; Frigg and Werndl 2012a). Whether a system spends most of its time in a $\delta$-prevalent macro-region depends on the dynamics. If, for instance, the dynamics is the identity function, states not initially in the $\delta$-prevalent macro-region will never evolve to the $\delta$-prevalent macro-region and spend most of their time there. Hence there will be no approach to equilibrium. We will come back to this point below in Section \ref{Existence Section}.\\

Another account \emph{defines equilibrium in terms of the Maxwell-Botzmann distribution}: a system is in equilibrium when its particles satisfy the Maxwell-Boltzmann distribution (equation \ref{rig}) (e.g., Penrose 1989). This is not a viable definition. The Maxwell-Boltzmann distribution is in fact the equilibrium distribution only for a limited class of systems, namely for systems consisting of particles in a finite container isolated from the environment with negligible interparticle forces. Examples 2 and 3 show that there are systems whose equilibrium distribution is not the Maxwell-Boltzmann distribution. In general, systems with non-negligible interactions will have equilibrium distributions that are different from the Maxwell-Boltzmann distribution (Gupta 2002). Defining equilibrium in terms of the Maxwell-Boltzmann distribution has therefore the false consequence that many systems of which we know that they are approaching equilibrium, will never approach equilibrium.\\

A third strategy justifies \emph{prevalence by maximum entropy considerations} along the following lines:\footnote{This strategy has been mentioned to us in conversation but is hard to track down in print. Albert's (2000) considerations concerning entropy seem to gesture in the direction of this third strategy.} we know from TD that, if left to itself, a system approaches equilibrium, and equilibrium is the maximum entropy state. Hence the Boltzmann entropy of a macro-state $S_{B}$ is maximal in equilibrium. Since $S_{B}$ is a monotonic function, the macro-state with the largest Boltzmann entropy is also the largest macro-state, which is the desired conclusion.\\

There are serious problems with the understanding of TD in this argument and with the implicit reductive claims. First, that a system, when left to itself, reaches equilibrium where the entropy is maximal is often regarded as a consequence of the Second Law of TD, but it is not. As Brown and Uffink point out (2001), that systems tend to approach equilibrium has to be added as an independent postulate, and they call this postulate the  `Minus First Law'. But the conclusion does not follow even if TD is amended with the Minus First Law. TD does not attribute an entropy to systems out of equilibrium at all. Thus, from a TD point of view characterising the approach to equilibrium as a process of entropy increase is meaningless!\\

Even if all these issues could be resolved, there would remain the question why the fact that the TD entropy reaches a maximum in equilibrium would imply that this also holds for the Boltzmann entropy. To justify this inference, the assumption would need to be made that the TD entropy reduces to the Boltzmann entropy. However, this is far from clear. A connection between the TD entropy and the Boltzmann entropy has been established only for ideal gases. Here the Sackur-Tatrode formula can be derived from BSM, and this shows that both entropies have the same functional dependence on thermodynamic state variables. Yet for systems with interactions no such results are known (cf.\ Frigg and Werndl 2011b). Also, there are well-known differences between the TD and the Boltzmann entropy. For example, the TD entropy is extensive but the Boltzmann entropy is not (Ainsworth 2012), and an extensive concept cannot reduce to a non-extensive concept (at least not without further qualifications).\\

One could try to get around these worries by saying that `equilibrium' is a primitive term of BSM and it is simply a definition that equilibrium is the macro-state with the largest macro-region. This is, however, would pull the rug from underneath every attempt to establish a connection between BSM and TD, which is too undesirable a conclusion to be entertained seriously.

\subsection{Defining Equilibrium: Formal Complications}\label{D-Formal}	

The standard conception also faces formidable formal problems. Even if one was willing to set aside (or simply ignore) the conceptual problems discussed in the last subsection and focus just on the calculations, the standard conception would not come out looking good. The main problem is that, at least in its current form, the standard conception makes assumptions that are so strong that the domain of application of the theory is in effect limited to dilute gases. This is far too narrow a scope for a theory that ought to provide a general explanation of equilibrium phenomena.\\

Justifications of the fact that equilibrium corresponds to the largest macro-region typically rely on the combinatorial argument (Example 1). But the combinatorial argument makes extremely strong assumptions. It assumes that the energy of a particle depends only on the cell in which it is located. This assumption applies, strictly speaking, \emph{only to systems with non-interacting particles, i.e. ideal gases} (Frigg 2008; Uffink 2007).\footnote{Strangely, the combinatorial argument does not deliver the correct conclusion even for ideal gases because the Maxwell-Boltzmann distribution does \emph{not} correspond to the equilibrium distribution for ideal gases. The reason why this argument fails for ideal gases will be discussed in Subsection~\ref{HT}.} Ideal gases are, perhaps, a good approximation for \emph{dilute gases}, i.e.\ gases of low density, and so the argument may deliver the approximately correct results for such systems. However, the argument remains silent about systems with stronger inter-particle forces such as liquids and solids. This is a serious limitation, and no suggestions have been made so far as to how it could be overcome.\\

One might try to circumvent the formal problems with the combinatorial argument by taking the Maxwell-Boltzmann distribution as ones definition of equilibrium and then trying to argue -- without appeal to combinatorial considerations -- that the part of the effective state space taken up by points with that distribution is large. This will everything but solve the problem (and even if it did, one would still be left with the conceptual issues mentioned above!). Maxwell's original 1860 derivation is in fact also dependent on the assumption of non-interaction. The assumption enters via the postulate that the probability distributions in different spatial directions can be factorised, which is true only if there is no interaction between particles (see Uffink 2007). Furthermore, the Maxwell-Boltzmann distribution \textit{by itself} implies nothing about the size of the corresponding macro-region. Arguments for the claim that the Maxwell-Boltzmann distribution corresponds to the largest macro-region appeal to the combinatorial argument. So we have come full circle. The conclusion is that disregarding conceptual problems provides no rescue: the calculations, at least in their current form, simply do not provide what is needed.

\subsection{The Approach to Equilibrium}\label{Approach}	

Let us now briefly turn to the second question. Since this question has been extensively discussed, we only offer a short summary of the main points and refer the reader to the relevant literature.\\

The currently most influential account in physics is the \emph{typicality account}. It originates in the work of Lebowitz (1993a, 1993b) and has been developed, among others, by Zangh\`i  (2005), Goldstein (2001) and Goldstein and Lebowitz (2004). The leading idea behind this account is that systems approach equilibrium because equilibrium micro-states are typical. It is our considered view this account is unsuccessful because it fails to take the system's dynamics into account (cf.\ Frigg 2009, 2010a; Frigg and Werndl 2012).\footnote{We criticize the idea that the approach to equilibrium takes place because micro-states are typical. Interpreting measures in BSM as typicality measures might still be fruitful (e.g., Werndl 2013).}\\

The canonical answer to Question 2 is given within the \emph{ergodic programme}. The leading idea is that systems approach equilibrium iff they are ergodic. To introduce the formal definition of ergodicity, we first need the notion of the long-run fraction of time a system spends in a region $A\in \Sigma_{Z}$:
\begin{eqnarray}\label{LF}
LF_{A}(x)=\lim_{t\rightarrow\infty}\frac{1}{t}\int_{0}^{t}1_{A}(T_{\tau}(x))d\tau\,\,\textnormal{for continuous time, i.e.}\,\,t\in\field{R},\,\,\\
LF_{A}(x)=\lim_{t\rightarrow\infty}\frac{1}{t}\sum_{\tau=0}^{t-1}1_{A}(T_{\tau}(x))\,\,\textnormal{for discrete time, i.e.}\,\,t\in\field{Z},\nonumber
\end{eqnarray}
\noindent where $1_{A}(x)$ is the characteristic function of $A$, i.e.\ $1_{A}(x)=1$ for $x\in A$ and $0$ otherwise. A measure-preserving dynamical system $(Z, \Sigma_{Z},\mu_{Z},T_{t})$ with a normalised measure $\mu_{Z}$ is \emph{ergodic} iff for any $A \in \Sigma_{Z}$:
\begin{equation}\label{ergodic}
LF_{A}(x)=\mu_{Z}(A),
\end{equation}
for all $x \in Z$ except for a set $W$ with $\mu_{Z}(W)=0$.\\

The ergodic approach has been criticised for a number of reasons, most notably for its inapplicability to realistic systems (for a summary of the criticisms see Frigg 2008, 121-126). In our (2011) offer a generalisation of this account based on the notion of epsilon-ergodicity and argue that dilute gases satisfy this condition. A system $(Z,\Sigma_{Z},\mu_{Z},T_{t})$ is epsilon-ergodic iff:\footnote{In detail: $(Z,\Sigma_{Z},\mu_{Z},T_{t})$ is \textit{$\varepsilon$-ergodic}, $\varepsilon\in\field{R},\,0\leq\varepsilon<1$, iff there is a set $\hat{Z}\subset Z$, $\mu_{Z}(\hat{Z})=1-\varepsilon$, with $T_{t}(\hat{Z})\subseteq\hat{Z}$ for all $t$, such that the system
$(\hat{Z},\Sigma_{\hat{Z}},\mu_{\hat{Z}},T_{t})$ is ergodic, where $\Sigma_{\hat{Z}}$ and $\mu_{\hat{Z}}$ is the $\sigma$-algebra $\Sigma_{Z}$ and the measure $\mu_{Z}$ restricted to $\hat{Z}$. A system $(Z,\Sigma_{Z},\mu_{Z},T_{t})$
is \textit{epsilon-ergodic} iff there exists a very small $\varepsilon$ for which the system is $\varepsilon$-ergodic.}
\begin{equation}\label{EE}
\textnormal{it is ergodic on a set}\,\,\hat{Z}\subseteq Z\,\,\textnormal{of measure}\,\,1-\varepsilon,\,\,\textnormal{where}\,\,\varepsilon\,\,\textnormal{is a very small real number}.
\end{equation}
Our arguments go some way to countering the inapplicability charge, but it remains silent about systems with stronger interactions such as fluids and liquids.\\

Finally, there is a family of proposals that grounds the approach to equilibrium in \emph{different kinds of probabilistic dynamics}. Boltzmann (1877) introduces the probability of a macro-state and postulates that this probability is proportional to its size. Since equilibrium is the largest state it is also the most likely state. Systems then evolve from less to more likely states, which explains the approach to equilibrium. Albert (2000) introduces conditional probabilities by conditionalising on the past state and exploiting the internal structure of macro-regions. The result of this account is that systems are overwhelemingly likely to approach up in equilibrium. These approaches are discussed in Frigg (2010b) and found wanting both for technical and conceptual reasons.\\

The conclusion we draw from the above is that there is no satisfactory general answer to Question 2.

\section{Redefining Equilibrium}\label{Redefining Equilibrium}

The failure of standard justificatory strategies prompts the search for an alternative solution. This solution, we submit, cannot be found by revising any of the approaches reviewed in the last section. We have to wipe the slate clean and start over. The leading idea of our approach is to reverse the direction of definition, as it were. While many previous approaches sought to define equilibrium in terms of micro-mechanical properties, we depart from a TD definition of equilibrium and then exploit the supervenience of macro-states on micro-states to `translate' this macro definition into micro language. The resulting definition of equilibrium is not only free from the conceptual and technical difficulties we have encountered in the last section; it also paves the ground for a general theorem (to which we turn in the next section) establishing that the equilibrium macro-region is the largest macro-region in a requisite sense.\\

As we have seen above, a system is in TD equilibrium if all change has ground to a halt and none of its thermodynamic properties vary with time. This state also has the character of an attractor: it is the state to which an isolated system converges when left alone and which it never leaves once it has got there. Furthermore, TD equilibrium is unique in the sense that the system always converges toward the \textit{same} equilibrium state. Bringing these points together one can give the following definition, which also incorporates the Minus First Law of TD:
\begin{quote}
\textit{Definition 1: TD Equilibrium.}
Consider an isolated system $S$ and a set of macro-states $M_{V_{1}, \ldots, V_{l}}$. If there is a macro-state $M_{V_{1}^{*}, ..., V_{l}^{*}}$ satisfying the following condition, then it is the equilibrium state of $S$: For \textit{all} initial states $M_{V_{1}, ..., V_{l}}$ there exists a time $t^{*}$ such that $M_{V_{1}, ..., V_{l}}(t)= M_{V_{1}^{*}, ..., V_{l}^{*}}$ for all $t \geq t^{*}$, where $M_{V_{1}, ..., V_{l}}(t)$ is the macro-state after $t$ time steps for a system that started initially in $M_{V_{1}, \ldots, V_{l}}$.\footnote{The time $t^{*}$ may depend on the initial state. This dependence can be avoided by changing the requirement to: there exists a time $t^*$ such that  for all initial states $M_{V_{1}, ..., V_{l}}$: $M_{V_{1}, ..., V_{l}}(t)= M_{V_{1}^{*}, ..., V_{l}^{*}}$ for all $t \geq t^{*}$.
 Nothing in what follows depends on this.}
\end{quote}

Exploiting the fact that macro-states supervene on micro-states this translates in the following definition of BSM equilibrium (the qualification `strict' will become clear soon):
\begin{quote}
\textit{Definition 2: Strict BSM Equilibrium.} Consider the same system $S$ as in Definition 1, described as measure-preserving deterministic system $(Z,\Sigma_{Z},\mu_{Z},T_{t})$ equipped with the macro-variables $\{v_1,\ldots,v_{l}\}$, and let $M(x)$ be the macro-state that supervenes on micro-state $x$. If there is a macro-state $M_{V_{1}^{*}, ..., V_{l}^{*}}$ satisfying the following condition, then it is the strict BSM equilibrium of $S$: For \textit{all} initial states $x\in Z$ there exists a time $t^{*}$ such that $M_{V_{1}, ..., V_{l}}(T_{t}(x))=M_{V_{1}^{*}, ..., V_{l}^{*}}$ for all $t \geq t^{*}$.
\end{quote}

Definition 2 is too rigid and there are two reasons for this. First, in SM, unlike in TD, we should not expect \textit{every} initial condition to approach equilibrium (e.g.\ Callender 2001). Indeed, it is reasonable to allow that there is a set of initial conditions of very small measure $\varepsilon$ which do not approach equilibrium.\\

Second, the systems under consideration exhibit Poincar\'{e} recurrence: as long as the `M' in SM refers to a mechanical theory that conserves state space volume (and there is widespread consensus about this),\footnote{Hamiltonian mechanics falls within this class, but the class is much broader.} any attempt to justify an approach to strict equilibrium in mechanical terms cannot succeed. The system will at some point return arbitrarily close to its initial condition, in violation of strict equilibrium (Frigg 2008; Uffink 2007). Furthermore, strict equilibrium is not only unattainable but also undesirable. Experimental results show that equilibrium is not the immutable state that classical TD presents us with because systems exhibit fluctuations away from equilibrium (MacDonald 1962; Wang et al.\ 2002). Thus strict equilibrium is actually \emph{unphysical} and adopting it would diminish the empirical adequacy of the theory.\\

To get around these difficulties, the condition that a system has to remain in equilibrium for all $t \geq t^{*}$ has to be relaxed. It is natural to postulate that the equilibrium state is the state in which the system spends most of the time in the long run. This can be done in two ways. The first is to demand that equilibrium is the state in which the system spends at least $\alpha$ of its time for $\alpha\in (\frac{1}{2}, \, 1]$. Recalling the notion of the long-run fraction of time a system spends in a region $A$ (equation \ref{LF}), we can state the first definition of equilibrium:
\begin{quote}
\textit{Definition 3: BSM $\alpha$-$\varepsilon$-Equilibrium.}
Consider the same system as in Definition 2. Let $\alpha$ be a real number in the interval $(\frac{1}{2}, \, 1]$, and let $\varepsilon$ be a very small positive real number. If there is a macro-state $M_{V_{1}^{*}, ..., V_{l}^{*}}$ satisfying the following condition, then it is the $\alpha$-$\varepsilon$-equilibrium state of $S$: There exists a set $Y\subseteq Z$ such that $\mu_{Z}(Y)\geq 1-\varepsilon$, and all initial states $x\in Y$ satisfy
\begin{equation}
LF_{Z_{M_{V_{1}^{*}, ..., V_{l}^{*}}}}\!(x) \, \geq \, \alpha.
\end{equation}
We then write $M_{\alpha\textnormal{-}\varepsilon\textnormal{-}eq}\, := \, M_{V_{1}^{*}, ..., V_{k}^{*}}$.\footnote{We assume that there are at least two macro-states of measure $\geq \varepsilon$ \label{epsilon} to avoid  that this definition can be trivially fulfilled (i.e., because there is only one macro-state or because macro-regions smaller than the largest macro-region -- regions that  correspond to states initially in non-equilibrium -- have a total measure smaller than $\varepsilon$ and hence it is irrelevant what happens to them).}
\end{quote}
In this definition $\alpha$ gives a lower bound for the fraction of time that the system spends in equilibrium.\footnote{Definition 3 is bears some similarity to Lavis' (2005, 255) notion of TD-likeness. However, unlike TD-likeness, Definition 3 makes no assumptions about the nature of fluctuations.} Intuitively one would like $\alpha$ to be close to one. However, nothing in the formal apparatus depends on this and so we do not build any such requirement into the theory. A determination of the correct value of $\alpha$ may depend on contextual factors and it is advantageous to keep options open.\\

The second way of relaxing the strict definition is to compare the time spent in different macro-states (Boltzmann 1877; Bricmont 2001). From such a comparative point of view it is natural to say that if there is a macro-state in which the system spends more time than in any other state, then that is the equilibrium state. This idea can be rendered precise as follows:
\begin{quote}
\textit{Definition 4: BSM $\gamma$-$\varepsilon$-Equilibrium.}
Consider the same system as in Definition 2. Let $\gamma$ be a real number in $(0, 1]$ and let $\varepsilon$ be a small positive real number. If there is a macro-state $M_{V_{1}^{*}, ..., V_{l}^{*}}$ satisfying the following condition, then it is the $\gamma$-$\varepsilon$ equilibrium state of $S$: There exists a set $Y\subseteq Z$ such that $\mu_{Z}(Y)\geq 1-\varepsilon$ and for all initial conditions $x\in Y$:
\begin{equation}
LF_{Z_{M_{V_{1}^{*}, ..., V_{l}^{*}}}}\!(x) \, \geq \,
LF_{Z_{M}}\!(x)+\gamma
\end{equation}
\noindent for all macro-states $M \neq M_{V_{1}^{*}, ..., V_{l}^{*}}$. We then write $M_{\gamma\textnormal{-}\varepsilon\textnormal{-}eq}\, := \, M_{V_{1}^{*}, ..., V_{k}^{*}}$.
\end{quote}

\noindent The parameter $\gamma$ gives a lower bound for the fraction of time that the system spends longer in $M_{\gamma\textnormal{-}\varepsilon\textnormal{-}eq}$ than in any other state. To facilitate language, we then say that the system spends \textit{$\gamma$-more-time} in $M_{\gamma\textnormal{-}\varepsilon\textnormal{-}eq}$. If, for instance, $\gamma=0.2$, then the fraction of time the system spends in $M_{\gamma\textnormal{-}\varepsilon\textnormal{-}eq}$ is at least 0.2 larger than the fraction of time it spends in any other macro-state.\\

As above, there is question about the correct value of the parameter, and intuitively one would like $\gamma$ to be as close to one as possible. However, again, nothing in the formal apparatus depends on $\gamma$ assuming specific values and so there is no need to enter into any commitments.\footnote{Both Definition 3 and 4 are time reversal invariant: the same equilibrium state would emerge under the time-reversed dynamics.}\\

The general proofs (in the next section) that the equilibrium state is the largest state will be based on these definitions. But before harvesting the fruits of our efforts, we would like to add a number of qualifications. First, it should be stressed that an important assumption in this characterisation of equilibrium is that $\mu_{Z}$ (and not some other measure) is the relevant measure.\\

Second, notice that an $\alpha$-$\varepsilon$-equilibrium is strictly stronger than a $\gamma$-$\varepsilon$-equilibrium. Whenever a system has an $\alpha$-$\varepsilon$-equilibrium, then it also has a $\gamma$-$\varepsilon$-equilibrium and $2\alpha-1$ provides a lower bound for $\gamma$. The converse need not hold: a system can have a $\gamma$-$\varepsilon$-equilibrium without having an $\alpha$-$\varepsilon$-equilibrium. Indeed this is the situation we encounter in the baker's gas with the macro-states $D_{bg}$ (Example 2) and the Kac-ring with the macro-states $M^{K}_{i}$ (Example 3). In both cases the largest macro-region corresponds to an $\gamma$-$\varepsilon$-equilibrium, but this region does not correspond to an $\alpha$-$\varepsilon$-equilibrium (Lavis 2005, 2008).The reason for this is that there are a vast number of non-equilibrium states. Thus while each non-equilibrium state is \textit{individually} much smaller than the equilibrium state (as per prevalence), \textit{taken together} all these non-equilibrium states can be of a considerable size. In fact, taken together, the non-equilibrium macro-states may take up a larger chunk than the equilibrium macro-state.\\

Third, we remain agnostic about issues of precedence. Both notions of equilibrium are legitimate and any preference for one over the other will depend on the context and purpose of the investigation. There is no non-arbitrary way to single out one as `true equilibrium'.\\

Fourth, that there is an approach to equilibrium is built into both definitions of equilibrium. If a state is not such that the system spends most of the time in it (in one of the two senses), then it simply is not an equilibrium state. Having an equilibrium state and there being an approach to equilibrium are really the two sides of the same coin. This avoids the ineptness of other approaches which have to make sense of systems where an equilibrium state exists but no approach to equilibrium takes place. The crucial question in our approach is: under what circumstances does a system have an equilibrium state at all. We turn to this question in Section~\ref{Existence Section}.\\

Fifth, it is part of the folklore of SM that the approach to equilibrium happens fairly quickly, and some may bemoan that this fact has not been built into the definition of equilibrium. There are good reasons to resist such a move. First, TD is completely silent about the speed at which processes take place (indeed, there is no parameter for time in the theory). Second, approaches to equilibrium happen at various speeds and not all are fast (e.g., large steel parts take months to cool down). For these reasons the approach to equilibrium being fast should not be part of a definition of equilibrium.\\

Sixth, some reductionists may feel that a definition of equilibrium in SM that is based on `top down translation' of its namesake in TD undermines the prospect of reducing TD to SM. They would argue that equilibrium has to be defined in purely mechanical terms, and must then be shown to agree with the TD definition of equilibrium. This point of view is not the only one and reduction can be had even if equilibrium is defined `top down'. For one, whether the above definition undercuts a reduction depends on the concept of reduction one entertains. For someone with a broadly Nagelian perspective on reduction there is no problem: the above definition provides a bridge law, which allows the derivation of the requisite macro regularities from the laws of the micro theory.  Similar arguments can be made in the framework of New Wave Reductionism (cf.\ Dizadji-Bahmani et al.\ 2010). Second, equilibrium is a macro concept: when describing a system as being in equilibrium, one looks at it in terms of macro-properties. From a micro point of view all there is are molecules bouncing around. They always bounce -- there is no such thing as a relaxation of particle motion to an immutable state. Hence the very notion of equilibrium is of questionable significance at the micro level, and a definition of equilibrium in macro terms is no heresy.

\section{General Proofs for $\beta$-Dominance and $\delta$-Prevalence}
The crucial question now is: are the $\alpha$-$\varepsilon$-equilibrium and $\gamma$-$\varepsilon$-equilibrium macro-regions large in a requisite sense? In this section we prove in full generality that this is so. And we emphasise that `proof' and `full generality' ought to be taken literally. Our argument is based on two mathematical theorems for which we provide rigorous mathematical proofs. The argument is fully general in that \textit{no} assumptions about the system's dynamics or the nature of the interaction between particles is made. The theorems apply to any isolated system no matter what its internal constitution.\\

Before stating the theorems we have to say what `large in a requisite sense' means, and unsurprisingly this will not be the same for the two notions of equilibrium. As we have seen in the last section, a system has an $\alpha$-$\varepsilon$-equilibrium if, in the long run, the trajectories starting in most initial conditions spend at least fraction $\alpha$ of their time in the equilibrium macro-region. A natural notion of largeness for this kind of equilibrium is that the equilibrium macro-region occupies approximately fraction $\alpha$ of the state space. This is tantamount to saying that this region is \emph{$\beta$-dominant}, where $\beta$ is roughly equal to $\alpha$. The case of the $\gamma$-$\varepsilon$-equilibrium is analogous. The same line of reasoning leads to the conclusion that the equilibrium macro-region being large means that it is $\delta$-prevalent for a value of $\delta$ that is roughly equal to $\gamma$.\\

It is worth pointing out that the definitions in the last section by no means prejudge that matter: neither Definition 3 nor Definition 4 make a statement about the relative size of the macro-regions $Z_{M_{\alpha\textnormal{-}\varepsilon\textnormal{-}eq}}$ or  $Z_{M_{\gamma\textnormal{-}\varepsilon\textnormal{-}eq}}$, nor do they in an obvious sense imply anything about it. Indeed, these regions being extremely small would be entirely compatible with the definitions. That these macroregions have the right size is established in the two theorems that we are stating now, and which we prove in the Appendix.

\begin{quote}
\emph{Dominance Theorem}: If  $M_{\alpha\textnormal{-}\varepsilon\textnormal{-}eq}$ is an $\alpha$-$\varepsilon$-equilibrium of system $S$, then $\mu_{Z}(Z_{M_{\alpha\textnormal{-}\varepsilon\textnormal{-}eq}}) \geq \beta$ for $\beta=\alpha(1-\varepsilon)$.\footnote{We assume that $\varepsilon$ is small enough so that $\alpha(1-\varepsilon)> \frac{1}{2}$.}
\end{quote}

\begin{quote}
\emph{Prevalence Theorem}: If $M_{\gamma\textnormal{-}\varepsilon\textnormal{-}eq}$ is a $\gamma$-$\varepsilon$-equilibrium of system $S$, then
$\mu_{Z}(Z_{M_{\gamma\textnormal{-}\varepsilon\textnormal{-}eq}}) \geq \mu_{Z}(Z_{M})+\gamma-\varepsilon$ for all macro-states $M\neq M_{\gamma\textnormal{-}\varepsilon\textnormal{-}eq}$.\footnote{We assume that $\varepsilon<\gamma$.}
\end{quote}

It is important to highlight that both theorems prove the conditional claim that \textit{if} there is an $\alpha$-$\varepsilon$-equilibrium/$\gamma$-$\varepsilon$-equilibrium, \textit{then} the corresponding equilibrium macro-region is dominant/prevalent. As with all conditionals, the crucial question is whether, and under what conditions, the antecedent holds. We turn to this issue now.

\section{The Existence of an Equilibrium State}\label{Existence Section}

In this section we address the vexed question under what conditions does an equilibrium state exist? On our view this question subsumes the question under what conditions the approach to equilibrium takes place. That a system approaches equilibrium is built into the notion of an equilibrium state. If a state is not such that the system spends most of the time in that state (in one of the two senses specified), then that state simply isn't an equilibrium state. In other words, if the system does not approach equilibrium, then there is no equilibrium. Having an equilibrium state and there being an approach to equilibrium are two sides of the same coin. So when we propose to inquire into the conditions under which an equilibrium state exists, we are not turning to a finicky question in mathematical physics. In fact, what is at stake is one of the core problems of SM, namely the approach to equilibrium.\\

The main message of this section is that for an equilibrium to exist \emph{three factors need to cooperate: the choice of macro-variables, the dynamics of the system, and the choice of the effective state space $Z$}. The cooperation between these factors can take different forms and there is more than one constellation that can lead to the existence of an equilibrium state. The important point is that the answer to the question of existence is holistic: it not only depends on three factors rather than one but also on the interplay between these factors. For these reasons we call these three factors the \textit{holist trinity}.\\

A number of previous proposals fail to appreciate this point. The problem of the approach to equilibrium has often been framed as the challenge to identify \textit{one} crucial property and show that the relevant systems possess this property. Ergodicity is a case in point. This is to start on the wrong foot. As we will see, ergodicity is neither necessary nor sufficient for the existence of an equilibrium state (which, again, incorporates the approach to equilibrium).\\

In the next subsection this trinity is introduced in an informal way and illustrated with examples. These examples show what requisite collaborations look like, and what can go wrong. In subsection \ref{Existence Theorem} we state a rigorous mathematical theorem (which we prove in the Appendix) providing necessary and sufficient conditions for the existence of an equilibrium state. In subsection \ref{Revisiting Erg} we revisit the ergodic account from the point of view proposed in this paper.

\subsection{The Holist Trinity}\label{HT}
\textit{\textbf{Macro-variables}.} The first condition is that the macrovariables must be the right ones. To illustrate this, consider the baker's gas (Example 2) with an odd number of particles and suppose that there is only one macro-variable $v_{1}$. This variable indicates whether more particles of the gas are on the right hand side of the container: it takes the value 1 if more particles are on the right hand side and the value 0 if this is not the case. This macro-variable leads to two macro-states with corresponding macro-regions $B_R$ and $B_{notR}$ with $\mu_B(B_{R})=1/2=\mu_B(B_{notR})$. (Note that $B_{notR}$  also contains those states in which an odd number of particles are exactly in the middle of the box and thus an equal number of particles is on the right hand side and on the left hand side of the box. Since these states make up a set of measure zero we have $\mu_B(B_{R})=1/2=\mu_B(B_{notR})$). Condition (\ref{ergodic}) of ergodicity is usually regarded as the dynamical condition most conducive to an approach to equilibrium, and the baker's gas is ergodic (Werndl 2009). Nevertheless there is no equilibrium in this case because, in the long run, the system spends half of its time in both macro-regions. By contrast, for the macro-states $D_{bg}$ there is a $\gamma$-$\varepsilon$-equilibrium for the very same dynamics. This illustrates that the existence of an equilibrium depends as much on the choice of macro-variables as it depends on the system's dynamical properties and that there is no dynamical condition that is `the right one' in absolute terms. \\

This also implies that if no macro-variables are considered at all, there can be no equilibrium. Obvious as this may seem, some confusion has resulted from ignoring this simple truism. Sklar (1973, 209) mounts an argument against the ergodic approach by pointing out that a system of two hard spheres in a box has the right dynamics (namely ergodicity) and yet fails to show an approach to equilibrium. It hardly comes as a surprise, though, that there is no approach to equilibrium if the system has no macro-variables associated with it in terms of which equilibrium could even be defined! A dynamical condition by itself just is not sufficient to guarantee an approach to equilibrium.\\

There is a difference in the macro-state structure of our three examples. The macro-states of dilute gases (Example 1) and of the baker's gas (Example 2) are \emph{local} in the sense that position matters: for micro-states which correspond to the same macro-state no differences in the (coarse-grained) position are allowed. Contrary to this, the macro-states of the Kac-ring (Example 3) are \emph{non-local} in the sense that the (coarse-grained) position does \emph{not} matter: there are micro-states where a site or a set of neighbouring sites has a different property but which correspond to the same macro-state.\footnote{There is also a similarity in the macro-state structure of the above examples. In all these examples macro-states are determined by combinatorial considerations of the single macro-space variable. That is, one considers a partition $\{\alpha_{1},\ldots,\alpha_{f}\}$, $f\geq 2$, of the state space of a single variable (particle or site), and then the macro-states are determined by counting the number of single variables taking a value in $\alpha_{1}$, in $\alpha_{2},\ldots$, in $\alpha_{f}$. Yet it is important to see that this is \emph{just one possible way} of defining macro-states. Other macro-states, such as the local macro-states for the lattice gas or the non-local macro-states for the gas mentioned in the next paragraph, are not determined in this way.}\\

Furthermore, this difference is \emph{not} necessitated by the systems under consideration. One could consider macro-states for the Kac-ring which are local. E.g., one could partition the pointers into cells of two neighbouring sites where there are three configurations (two black, two white, one black and one white) and then define macro-states according to the different configurations of all the cells. Indeed, sometimes one will want to take clustering into account, and then macro-states different from $M^{K}_{i}$ will need to be considered. Conversely, one could consider macro-states for dilute gases which are non-local, e.g., by requiring, on top of the classical macro-state structure, that one obtains the same macro-state if the particles in a box are rotated in space along one of the three axes or along a diagonal.\footnote{Here it should be added that sometimes one will find that the macro-variables previously considered are too fine. As a consequence, one turns to new macro-variables which are defined by the previously considered macro-variables taking values in a certain \emph{range or interval}. To provide an example, for the Kac-ring, particularly when the number of sites is very large, one may not want to consider the total number of up spins but different ranges of the total number of up spins (e.g.\ as in Lavis 2008). Similarly, for dilute gases when the number of particles is large one might not want to consider the total numbers of particles in cell $\omega^{dg}_{1}$, $\omega^{dg}_{2}$ etc., but different ranges of the total number of particles in cell $\omega_{1}^{dg}$, $\omega_{2}^{dg}$ etc.}\\

These observations highlight that some care is needed in choosing one's macro-variables. Different choices are possible, and these choices lead to different conclusions about the equilibrium behaviour of the system. We do not think that there is right and wrong in this choice. What macro-variables one picks in a given situation will depend on the macroscopic properties of interest, which can, or course, vary depending on the context of the investigation and on the situation at hand.\\

\noindent\textit{\textbf{Dynamics}.} The existence of an equilibrium depends as much on the dynamics of the system as it depends on the choice of macro-variables. Whatever the macro-variables, if the dynamics is not the right one, then there will be no approach to equilibrium. Hence, the converse of the Dominance and Prevalence Theorems is \textit{not} true. That is, the following conditional is false: if there is a $\beta$-dominant/$\delta$-prevalent macro-region, then this macro-region corresponds to a $\alpha$-$\varepsilon$-equilibrium/$\gamma$-$\varepsilon$-equilibrium.\\

Let us give a few simple examples to drive this point home: suppose that the dynamics is the identity function. Then there can be no approach to equilibrium because states in a small macro-region will always stay in this region. Similarly, a Kac-ring without spin flippers has no equilibrium, and a system of uncoupled harmonic oscillators with the ideal gas macro-state structure has no equilibrium.
To provide a more general example, suppose that the dynamics is such that states initially in the largest macro-region always remain in the largest macro-region and states initially in smaller macro-regions only evolve into states in these smaller macro-regions. Then there can be no approach to equilibrium because non-equilibrium states will not evolve into equilibrium.\\

\noindent \textit{\textbf{Identifying the correct $Z$}.}
A number of considerations in connection with equilibrium depend on the choice of $Z$, which is the set relative to which macro-regions are defined. Intuitively speaking, $Z$ are subsets whose states evolve to the same equilibrium macro-state. Hence, crucially, the very existence of an equilibrium state depends on the correct choice of $Z$. There can be situations where a system has an equilibrium with respect to one choice of $Z$ but not with respect another choice of $Z$. Most importantly, if there is an equilibrium relative to some set $Z$, this does not imply that there exists an equilibrium on a superset of this set.\\

Let us put this observation into perspective. The `choice of $Z$ problem' can be subsumed under the question whether the combination of macro-variables and dynamics is the right one in the following sense: even if there is an equilibrium (in one of two senses) relative to a (correctly chosen) set $Z$, whenever $Z$ is chosen wrongly, then for the given macro-state structure and dynamics there will be no equilibrium. In particular, relative to given macro-state structure and dynamics, there are choices of $Z$ so that there is an equilibrium relative to $Z$ even though there does not exist an equilibrium for a superset of $Z$.\\

For this reason determining the equilibrium state as the macro-region of largest measure will only work if the correct set is chosen. Most importantly, if there is an equilibrium relative to a set $Z$, but one instead chooses a superset for the maximation procedure, one cannot expect to arrive at the correct result. In this case there might be a macro-region of largest measure, but it does not correspond to equilibrium because there is no equilibrium relative to the superset.\\

Let us illustrate this with two examples. First, consider a dilute gas with the distributions $D_{dg}$ as macro-states (Example 1). Suppose that one decides to identify the equilibrium macro-state of $\Gamma$ by determining the largest macro-region of $\Gamma$. Clearly, the result one obtains is that the largest macro-region corresponds to the uniform distribution. However, we know that equilibrium is given by the Maxwell-Boltzmann distribution (\ref{rig}) and \emph{not} the uniform distribution! What has gone wrong? The answer is that \emph{relative to $\Gamma$ no equilibrium exists} because there are different equilibria for different total energies of the system (as reflected by the Maxwell-Boltzmann distribution, which depends on the total energy $E$). That is, an equilibrium only exists relative to the set $\Gamma_{E}$ but not relative to $\Gamma$. Note that the reason that one does not obtain the correct equilibrium distribution by determining the largest macro-region of $\Gamma$ is not that $\Gamma$ is decomposable under the dynamics (e.g., the set $K$ of the KAC-ring is decomposable, but it still has a unique equilibrium – cf.\ the end of Subsection 6.3). Rather, the reason is that the equilibrium macro-state is dependent on the energy value and thus the set $Z$ relative to which equilibrium is defined cannot be $\Gamma$.\\

As a second example consider an ideal gas consisting of $N$ particles with mass $m$ moving on a three-dimensional torus. Suppose that one decides to determine the equilibrium macro-state of $\Gamma_{E}$ by following the standard procedure of the combinatorial argument. Hence one calculates the largest macro-region of $\Gamma_{E}$ and arrives at the Maxwell-Boltzmann distribution (\ref{rig}). However, for an ideal gas the momentum $p_{i}$ of each particle is a constants of motion and $e_i=p_i^{2}/(2m)$. Therefore, if the ideal gas with a certain energy level $E=\sum_{i=1}^{N}e_{i}$ starts in a micro-state where the momenta of the particles are not distributed according to the Maxwell-Boltzmann distribution, they will never be distributed according to the Maxwell-Boltzmann distribution. Hence for an ideal gas the Maxwell-Boltzmann distribution does \emph{not} correspond to equilibrium. The combinatorial argument does \emph{not} deliver the correct conclusion for ideal gases because the largest macro-region is determined relative to the set $\Gamma_{E}$ where no equilibrium exists! There is a $\gamma-\varepsilon$-equilibrium (namely, where the particles are uniformly distributed in space with constant $p$), but it is relative to the hypersurface of constant momentum $Z_{p}$.\\

Summing up, we have shown that the existence of an equilibrium depends on the harmonious interplay of three factors: the choice of macro-states, the dynamics of the system and the choice of the effective state space $Z$. The challenge now is to say in precise terms what `harmonious' amounts to. The Existence Theorem in the next subsection conclusively answers this question.

\subsection{The Existence Theorem }\label{Existence Theorem}

In this subsection we introduce a theorem providing necessary and sufficient conditions for the existence of an equilibrium state. Like the theorems we have seen earlier, it is fully general in that it makes no assumptions about the system's dynamics other than that it be measure-preserving. \\

Before stating the theorem (the proof is given in the Appendix), we have to introduce the \textit{Ergodic Decomposition Theorem}, which is a crucial theorem in ergodic theory (cf.\ Petersen 1983, 81). An ergodic decomposition of a system is a partition of the state space into cells so that the cells are invariant under the dyanamics (i.e. are mapped onto themselves) and that, for any arbitrary cell, the dynamics, when restricted to a cell, is ergodic (it is allowed that the regions are of measure zero and that there are uncountably many of them). More colloquially: there is an ergodic decomposition if one can slice up the state space in different parts in each of which the dynamics is ergodic. The Ergodic Decomposition Theorem makes the – surprising – statement that such a decomposition exists for \textit{every} measure-preserving dynamical system with a normalized measure, and that the decomposition is unique. In other words, the dynamics of a system can be as complex as we like and the interactions between the constituents of the system can be as strong and intricate as we like, and yet there exists a unique ergodic decomposition of the state space of the system. As an example consider the Kac-ring: the system is not ergodic on the entire state space, but it does have an ergodic decomposition (Lavis 2005). Another simple example is the harmonic oscillator, whose state space can be decomposed into ellipses on which the motion is ergodic.\\

 For what follows it is helpful to have a more formal rendering of an ergodic decomposition (the precise formulation of the Ergodic Decomposition Theorem can be found in the Appendix). Consider the system $(Z,\Sigma_{Z},\mu_{Z},T_{t})$. Let $\Omega$ be an index set, which can but need not be countable. Let $Z_{\omega}$, $\omega \in \Omega$, be the cells into which the system's state space can be decomposed.  $\Omega$ comes equipped with a a probability measure $\nu$, which tells one
 how large a set of $Z_{\omega}$ is that are characterised by certain values of $\omega$. Furthermore, let $T_{t}$ be the restriction of the dynamics of the system to $Z_{\omega}$, and let
  $\Sigma_{\omega}$ and $\mu_{\omega}$, respectively, be the sigma algebra and measure defined on $Z_{\omega}$. These can be gathered together in  `components' $C_{\omega}=(Z_{\omega}, \Sigma_{\omega}, \mu_{\omega}, T_{t})$. The Ergodic Decomposition Theorem says that for every system $(Z,\Sigma_{Z},\mu_{Z},T_{t})$ there exists a unique set of ergodic $C_{\omega}$ so that the system itself amounts to the collection of all the $C_{\omega}$.\\

We are now in a position to state the core result:
\begin{quote}
\noindent \emph{\textbf{Existence Theorem}}:
Consider a measure-preserving system $(Z,\Sigma_{Z},\mu_{Z},T_{t})$ with macro-regions $Z_{M_{V_{1}, \ldots,V_{l}}}$ and let $C_{\omega}=(Z_{\omega},\Sigma_{\omega},\mu_{\omega},T_{t})$, $\omega \in \Omega$, be its ergodic decomposition. Then the following two biconditionals are true:\\\\
\noindent \textit{$\alpha$-$\varepsilon$-equilibrium}:
\noindent There exists an $\alpha$-$\varepsilon$-equilibrium iff there is a macro-state $\hat{M}$ such that for every $C_{\omega}$:
\begin{equation}\label{ggg}
\mu_{\omega}(Z_{\omega}\!\cap\! Z_{\hat{M}}) \, \geq \, \alpha,
\end{equation}
except for components $C_{\omega}$ with $\omega\in\Omega'$, $\mu_{Z}(\cup_{\omega\in\Omega'}Z_{\omega})\leq\varepsilon$. $\hat{M}$ is then the $\alpha$-$\varepsilon$-equilibrium state.\\\\
\noindent \textit{$\gamma$-$\varepsilon$-equilibrium}:
\noindent There exists a $\gamma$-$\varepsilon$-equilibrium iff there is a macro-state $\hat{M}$ such that for every $C_{\omega} $ and any $M \neq \hat{M}$
\begin{equation}\label{stress}
\mu_{\omega}(Z_{\omega}\!\cap\! Z_{\hat{M}}) \, \geq \, \mu_{\omega}(Z_{\omega}\!\cap\! Z_{M})+\gamma,
\end{equation}
except for components $C_{\omega}$ with $\omega\in\Omega'$, $\mu_{Z}(\cup_{\omega\in\Omega'}Z_{\omega})\leq\varepsilon$. $\hat{M}$ is then the $\gamma$-$\varepsilon$-equilibrium state.
\end{quote}

\noindent Intuitively, the theorems say that there is an $\alpha$-$\varepsilon$-equilibrium/$\gamma$-$\varepsilon$-equilibrium iff if the system's state space is split up into invariant regions on which the motion is ergodic and the equilibrium macro-state takes up at least $\alpha$ of each region/the equilibrium region is larger than any other macro-region, except for regions of total measure $\varepsilon$. If we have found a space that matches these conditions, then it plays the role of the effective state space $Z$.\\

It is important to note that there may be many different macro-state/dynamics/$Z$ triplets that make the Existence Theorem true.
The Theorem gives the foundation for a research programme aiming to find and classify all these triplets. We will classify these triplets for the three examples we have introduced soon (at the end of Section \ref{Revisiting Erg}). Before we do so, it will be helpful to comment on the ergodic approach because it can be interpreted as a special case of the existence theorem.

\subsection{Revisiting the Ergodic Account}\label{Revisiting Erg}
The canonical explanation of equilibrium behaviour is given with the ergodic approach. We now revisit this approach and show that it can be interpreted \emph{as an instance of the Existence Theorem, i.e.\ as providing a triplet that satisfies the above conditions}. \\

It is often claimed that the approach to equilibrium can be explained with the dynamical condition of ergodicity (equation \ref{ergodic}) or epsilon-ergodicity (equation \ref{EE}). The results of this paper clarify these claims. First, as pointed out in the previous subsection, if the macro-variables are not the right ones, then even ergodicity or epsilon-ergodicity will \emph{not} imply that the approach to equilibrium takes place. However, second, proponents of ergodicity and epsilon-ergodicity as an explanation of the approach to equilibrium often assume that there is a $\beta$-prevalent macro-region/an $\alpha$-dominant macro-region (e.g.\ Frigg and Werndl 2011a, 2012b). Then this indeed leads to particularly simple cases of the Existence Theorem, implying that the macro-region corresponds to an $\alpha$-$\varepsilon$-equilibrium/a $\gamma$-$\varepsilon$-equilibrium. More specifically, the following two corollaries hold (proofs are given in the Appendix):
\begin{quote}
\noindent\emph{Ergodicity-Corollary}:
 Suppose that the measure-preserving system $(Z,\Sigma_{Z},\mu_{Z},T_{t})$ is ergodic. Then the following are true:
(a) If the system has a macro-region $Z_{\hat{M}}$ that is $\beta$-dominant, $\hat{M}$ is an $\alpha$-$\varepsilon$-equilibrium for $\alpha=\beta$.
(b) If the system has a macro-region $Z_{\hat{M}}$ that is $\delta$-prevalent, $\hat{M}$ is a $\gamma$-$\varepsilon$-equilibrium for $\gamma=\delta$.
\end{quote}

\begin{quote}
\noindent\emph{Epsilon-Ergodicity-Corollary}:
\noindent Suppose that the measure-preserving system $(Z,\Sigma_{Z},\mu_{Z},T_{t})$ is epsilon-ergodic. Then the following are true:
(a) If the system has a macro-region $Z_{\hat{M}}$ that is $\beta$-dominant for $\beta-\varepsilon>\frac{1}{2}$, $Z_{\hat{M}}$ is a $\alpha$-$\varepsilon$-equilibrium for $\alpha=\beta-\varepsilon$.
(b) If the system has a macro-region $Z_{\hat{M}}$ that is $\delta$-prevalent for $\delta-\varepsilon>0$, $Z_{\hat{M}}$ is a $\gamma$-$\varepsilon$-equilibrium for $\gamma = \delta-\varepsilon$.
\end{quote}
Yet it is important to keep in mind that ergodicity and epsilon-ergodicity are just examples of dynamical conditions for which an equilibrium exists. As shown by the Existence Theorem, the dynamics need not be ergodic or epsilon-ergodic for there to be an equilibrium. \\

Let us now come back to classify the triplets for the examples introduced in Section~\ref{BSM}, which will at the same time  illustrate the role of the ergodic approach.  First of all, consider Example 1 of the dilute gas (with the macro-state structure of the combinatorial argument), which, from experience, is regarded to have a $\alpha$-$\varepsilon$ equilibrium (and thus also a $\gamma$-$\varepsilon$-equilibrium). From a mathematical perspective the dynamics relative to $\Gamma_{E}$ is not well understood. Frigg and Werndl (2011a, 2012b) have argued that gases are epsilon-ergodic, and then this would be an instance of the Epsilon-Ergodicity-Corollary.  This is plausible, but it is not the only possibility. The Existence Theorem tells us what we know for sure. Namely: in order for the system to have a $\alpha$-$\varepsilon$-equilibrium/a $\gamma$-$\varepsilon$-equilibrium, whatever the dynamics, it has to be such that on each ergodic region the measure of the equilibrium macro-region is at least $\alpha$/the equilibrium macro-region is larger (by at least $\gamma$) than any other macro-region (except for regions of measure $\varepsilon$). Note that for the example of the dilute gas it is crucial that the right set $Z$ is chosen. The existence theorem is only satisfied relative to $Z=\Gamma_{E}$ but not relative to $Z=\Gamma$. If it were satisfied for $Z=\Gamma$, we would find that the solutions spend most of their time in the uniform distribution where $N_i=N/l$ (the macro-state with the largest measure relative to $\Gamma$), which is clearly not what we find. This is as it should be because there is no unique equilibrium for all states in $\Gamma$, but there are several different equilibria for different energy values.\\

Let us also briefly comment on the ideal gas on a torus with the macro-state structure as in the combinatorial argument. For $Z=\Gamma_{p}$ (the hypersurface determined by the constant momenta of the particles) it can be shown that the motion is ergodic (cf.\ Lavis 2005). Because the uniform distribution takes up more than any other macro-region on $\Gamma_{p}$, the Ergodicity Corollary implies that the uniform distribution of the particles corresponds to a $\gamma$-$\varepsilon$-equilibrium. Note that there is no $\alpha$-$\varepsilon$-equilibrium because none of the macro-regions is larger than measure $1/2$ (cf.\ Lavis 2005, 2008). For the ideal gas it is again crucial that the right set $Z$ is chosen. If instead $Z=\Gamma_{E}$ were considered, the Existence Theorem would rightly tell us that there is no equilibrium: For $Z=\Gamma_{E}$  the macro-region of largest size is the one corresponding to the Maxwell-Boltzmann distribution (\ref{rig}). But this distribution cannot correspond to equilibrium because the momenta of the particles of an ideal gas are constant. Hence if the momenta are not initially distributed as required by the Maxwell-Boltzmann distribution, they will never be distributed in this way.\\

Let us now turn to Example 2 of the baker's gas with the macro-states $D_{bg}$. Here there is an $\gamma$-$\varepsilon$-equilibrium  because there is a macro-region larger than any other macro-region and the dynamics is ergodic (Ergodicity-Corollary) (cf.\ Lavis 2005). Note that there is no $\alpha$-$\varepsilon$ equilibrium because all macro-regions are smaller than $1/2$ (cf.\ Lavis 2005).\\

Finally, Example 3 of the Kac-ring with the macro-states $M^{K}_{i}$ is \emph{not} ergodic. Yet there is still an $\gamma$-$\varepsilon$-equilibrium because the system decomposes into ergodic components and the equilibrium macro-region is larger than any other macro-region on each ergodic component (except for regions of measure $\varepsilon$) (Bricmont 1995; Lavis 2005). It is interesting that the motion of the Kac-ring is periodic, illustrating that an approach to equilibrium is also compatible with periodic motion. Note, again, there there is no $\alpha$-$\varepsilon$ equilibrium because all macro-regions are smaller than $1/2$ (cf.\ Lavis 2005).

\section{Conclusion}

What justifies the association of equilibrium with the largest macro-region in Boltzmannian statistical mechanics? We reviewed currently available answers to this question and found them wanting both for conceptual and technical reasons. We proposed a new conception of equilibrium and proved a mathematical theorem which establishes in full generality that if there is an $\alpha$-$\varepsilon$-equilibrium/$\gamma$-$\varepsilon$-equilibrium, then the corresponding equilibrium macro-region is $\beta$-dominant/$\delta$-prevalent. We then turned to the question of the approach to equilibrium, on which there exists no satisfactory general answer so far. In our account, this question is replaced by the question when an equilibrium state exists. We proved the (again fully general) Existence Theorem, providing necessary and sufficient conditions for the existence of an equilibrium state. This theorem re-orientates the discussion about equilibrium, which should focus on finding triplets of macro-variables, dynamical conditions, and effective state spaces that satisfy the conditions of the theorem. There are many triplets that satisfy the conditions of the Existence Theorem. Finding and describing at least some of them is a new research programme in the foundation of statistical mechanics.

\pagebreak

\section{Appendix}

\subsection{Proof of the Dominance Theorem}\label{A2}
The proof appeals to the powerful \emph{Ergodic Decomposition Theorem} (cf.\ Petersen 1983, 81), stating that for a measure-preserving deterministic system $(Z,\Sigma_{Z},\mu_{Z},T_{t})$ the set $Z$ is the disjoint union of sets $Z_{\omega}$, each equipped with a $\sigma$-algebra $\Sigma_{Z_{\omega}}$ and a probability measure $\mu_{\omega}$, and $T_{t}$ is ergodic on each $(Z_{\omega},\Sigma_{Z_{\omega}},\mu_{\omega})$. The indexing set is also a probability space $(\Omega,\Sigma_{\Omega},P)$, and for any square integrable function $f$ it holds that:
\begin{equation}
\int_{Z}fd\mu_{Z}=\int_{\Omega}\int_{Z_{\omega}}fd\mu_{\omega}dP.
\end{equation}

Suppose that the system has an $\alpha$-$\varepsilon$-equilibrium $M_{\alpha\textnormal{-}\varepsilon\textnormal{-}eq}$. Application of the Ergodic Decomposition Theorem for $f=1_{Z_{M}}(x)$, where $M$ is a an arbitrary macro-state, yields:
\begin{equation}\label{muede}
\mu_{Z}(Z_{M})=\int_{Z}1_{Z_{M}}(x)d\mu_{Z}=
\int_{\Omega}\int_{Z_{\omega}}1_{Z_{M}}(x)d\mu_{\omega}dP.
\end{equation}

For an ergodic system $(Z_{\omega},\Sigma_{Z_{\omega}},\mu_{\omega},T_{t})$ the long-run time average $LF_{Z_{M}}(x)$ (equation~(\ref{LF})) equals the measure of $Z_{M}\cap Z_{\omega}$ (cf.\ equation \ref{ergodic}). Hence for almost all $x\in Z_{\omega}$:
\begin{equation}
LF_{Z_{M}}(x)=
\int_{Z_{\omega}}1_{Z_{M}}(x)d\mu_{\omega}=\mu_{\omega}(Z_{M}\cap Z_{\omega}).
\end{equation}

From the definition of an $\alpha$-$\varepsilon$-equilibrium and because $T_{t}$ acts ergodically on each $(Z_{\omega},\Sigma_{Z_{\omega}},\mu_{\omega})$, for almost all
$x\in Z_{\omega}$, $Z_{\omega}\subseteq Y$:
\begin{equation}\label{WWW}
\alpha\leq LF_{Z_{M_{\alpha\textnormal{-}\varepsilon\textnormal{-}eq}}}(x)=
\int_{Z_{\omega}}1_{Z_{M_{\alpha\textnormal{-}\varepsilon\textnormal{-}eq}}}(x)d\mu_{\omega}.
\end{equation}

A comment is in order here about why we can assume in the statement before equation (\ref{WWW}) that $Z_{\omega}\subseteq Y$. Since the motion restricted to a given $Z_{\omega}$ is ergodic, either for all points in $Z_{\omega}$ the long-run fraction of time the system spends in the equilibrium region is greater than $\alpha$ (in which case $Z_{\omega}\subseteq Y$), or for all points in $Z_{\omega}$ the long-run fraction of time the system spends in the equilibrium region is not greater than $\alpha$ (in which case $Z_{\omega}\subseteq Z\setminus Y$). Hence the behaviour of states in $Y$ can be analysed by considering all the $Z_{\omega}$ which are a subset of $Y$ and all the other $Z_{\omega}$ can simply be set aside. \\

Now suppose for a moment that $\mu_{Z}(Y)=1$. Then from equation~(\ref{muede}):
\begin{equation}
\alpha=\int_{\Omega}\alpha dP\leq \mu_{Z}(Z_{M_{\alpha\textnormal{-}\varepsilon\textnormal{-}eq}}).
\end{equation}
Hence if $\mu_{Z}(Y)=1-\varepsilon$, it follows from equation~(\ref{muede}) that:
\begin{equation}
\alpha(1-\varepsilon)\leq \mu_{Z}(Z_{M_{\alpha\textnormal{-}\varepsilon\textnormal{-}eq}}).
\end{equation}

\subsection{Proof of the Prevalence Theorem}\label{A333}
The proof of the Prevalence Theorem proceeds similarly. From the definition of an $\gamma$-$\varepsilon$-equilibrium $M_{\gamma\textnormal{-}\varepsilon\textnormal{-}eq}$ and because $T_{t}$ acts ergodically on each $(Z_{\omega},\Sigma_{Z_{\omega}},\mu_{\omega})$:
\begin{equation}
\int_{Z_{\omega}}1_{Z_{M}}(x)d\mu_{\omega} +\gamma
=LF_{Z_{M}}(x)+\gamma\leq LF_{Z_{M_{\gamma\textnormal{-}\varepsilon\textnormal{-}eq}}}(x)=
\int_{Z_{\omega}}1_{Z_{M_{\varepsilon\textnormal{-}eq}}}(x)d\mu_{\omega},
\end{equation}
for almost all
$x\in Z_{\omega}$, $Z_{\omega}\subseteq Y$ and all macro-states $M$ with $M\neq M_{\gamma\textnormal{-}\varepsilon\textnormal{-}eq}$. Note that, as for the proof of the dominance theorem, the behaviour of states in $Y$ can be analysed by considering all the $Z_{\omega}$ with $Z_{\omega}\subset Y$ (because either  $Z_{\omega}\subseteq Y$ or $Z_{\omega}\subseteq Z\setminus Y$). \\

Suppose for a moment that $\mu_{Z}(Y)=1$. Then from equation~(\ref{muede}):
\begin{equation}
\mu_{Z}(Z_{M})+\gamma\leq \mu_{Z}(Z_{M_{\gamma\textnormal{-}\varepsilon\textnormal{-}eq}}),
\end{equation}
for all macro-states $M$ with $M\neq M_{\gamma\textnormal{-}\varepsilon\textnormal{-}eq}$. \\

Hence if $\mu_{Z}(Y)=1-\varepsilon$, it follows from equation~(\ref{muede}) that for all $M\neq M_{\gamma\textnormal{-}\varepsilon\textnormal{-}eq}$:
\begin{equation}
\mu_{Z}(Z_{M})+\gamma-\varepsilon\leq\mu_{Z}(Z_{M_{\gamma\textnormal{-}\varepsilon\textnormal{-}eq}}),
\end{equation}
for all macro-states $M$ with $M\neq M_{\gamma\textnormal{-}\varepsilon\textnormal{-}eq}$.

\subsection{Proof of the Existence Theorem}\label{PDT}
Let us first consider the case of an $\alpha$-$\varepsilon$-equilibrium.\\

$\Rightarrow$:  Assume that there exists an $\alpha$-$\varepsilon$-equilibrium $M_{\alpha\textnormal{-}\varepsilon\textnormal{-}eq}$. We again appeal to the Ergodic Decomposition Theorem and consider the decomposition of the system into ergodic components $C_{\omega}=(Z_{\omega},\Sigma_{Z_{\omega}},\mu_{\omega},T_{t})$ (cf.\ Subsection~\ref{A2}).

The definition of an $\alpha$-$\varepsilon$-equilibrium and equation (\ref{muede}) imply that
there exists a $\hat{M}$, namely $\hat{M}= M_{\alpha\textnormal{-}\varepsilon\textnormal{-}eq}$, such that the following holds:
for almost all $x\in Z_{\omega}$ and for all components $C_{\omega}$ except, maybe, for components $C_{\omega}$, $\omega\in\Omega'$, with $\mu_{Z}(\cup_{\omega\in\Omega'}Z_{\omega}) \leq \varepsilon$:
\begin{equation}
\alpha\leq LF_{Z_{\hat{M}}}(x)=
\int_{Z_{\omega}}1_{Z_{\hat{M}}}(x)d\mu_{\omega}=
\mu_{\omega}(Z_{\hat{M}}\cap Z_{\omega}),
\end{equation} which is the desired claim.\\

\noindent $\Leftarrow$: Conversely, suppose that there exists a macro-state $\hat{M}$ such that for the ergodic decomposition into components $C_{\omega}=(Z_{\omega},\Sigma_{Z_{\omega}},\mu_{\omega},T_{t})$ it holds that:
\begin{equation}
\mu_{\omega}(Z_{\omega}\!\cap\! Z_{\hat{M}})\geq\alpha \label{crazy2}
\end{equation}
except for components $C_{\omega}$ with $\omega\in\Omega'$, $\mu_{Z}(\cup_{\omega\in\Omega'}Z_{\omega})\leq\varepsilon$.\\

It follows from condition (\ref{crazy2}) that for all $x\in Y$ with $Y=Z\setminus(\cup_{\omega\in\Omega'}Z_{\omega})$, $\mu_{Z}(Y)=1-\varepsilon$:
\begin{equation}
\alpha \leq \mu_{\omega}(Z_{\hat{M}}\cap Z_{\omega})=\int_{Z_{\omega}}1_{Z_{\hat{M}}}(x)d\mu_{\omega}=LF_{Z_{\hat{M}}}(x),
\end{equation}
which means that $\hat{M}$ fulfills the definition of an $\alpha$-$\varepsilon$-equilibrium.\\

\noindent Let us now turn to the case of a $\gamma$-$\varepsilon$-equilibrium. \\

$\Rightarrow$: Assume that there exists a $\gamma$-$\varepsilon$-equilibrium $M_{\gamma\textnormal{-}\varepsilon\textnormal{-}eq}$ and consider the ergodic decomposition of the system into components $C_{\omega}=(Z_{\omega},\Sigma_{Z_{\omega}},\mu_{\omega},T_{t})$. Then the definition of a $\gamma$-$\varepsilon$-equilibrium and equation (\ref{muede}) imply that there is a $\hat{M}$, namely $\hat{M}=M_{\gamma\textnormal{-}\varepsilon\textnormal{-}eq}$ such that the following holds:
for almost all $x\in Z_{\omega}$ and for all components $C_{\omega}$ except, maybe, for components $C_{\omega}$, $\omega\in\Omega'$, with $\mu_{Z}(\cup_{\omega\in\Omega'}Z_{\omega}) \leq \varepsilon$:
\begin{eqnarray}
\mu_{\omega}(Z_{M}\cap Z_{\omega})+\gamma=\int_{Z_{\omega}}1_{Z_{M}}(x)d\mu_{\omega}+\gamma=LF_{Z_{M}}(x)+\gamma\\
\leq LF_{Z_{\hat{M}}}(x)=
\int_{Z_{\omega}}1_{Z_{\hat{M}}}(x)d\mu_{\omega}=\mu_{\omega}(Z_{\hat{M}}\cap Z_{\omega}),
\end{eqnarray}
for all macro-states $M$ with $M\neq \hat{M}$, which is the desired claim.\\

\noindent $\Leftarrow$: Conversely, suppose that there exists a macro-state $\hat{M}$ such that for the ergodic decomposition of the system  into components $C_{\omega}=(Z_{\omega},\Sigma_{Z_{\omega}},\mu_{\omega},T_{t})$ it holds that:
\begin{eqnarray}
\mu_{\omega}(Z_{\omega}\cap Z_{\hat{M}})&\geq& \mu_{\omega}(Z_{\omega}\cap Z_{M})+\gamma\,\,\,\textnormal{for}\,\,\,\textnormal{any}\,\,\,M\neq \hat{M}, \label{crazy1}
\end{eqnarray}
except for components $C_{\omega}$ with $\omega\in\Omega'$, $\mu_{Z}(\cup_{\omega\in\Omega'}Z_{\omega})\leq\varepsilon$.\\

Condition (\ref{crazy1}) implies that for all $x\in Y$ with $Y=Z\setminus(\cup_{\omega\in\Omega'}Z_{\omega})$, $\mu_{Z}(Y)=1-\varepsilon$:
\begin{eqnarray}
LF_{Z_{M}}(x)+\gamma=\int_{Z_{\omega}}1_{Z_{M}}(x)d\mu_{\omega}+\gamma=\mu_{\omega}(Z_{M}\cap Z_{\omega})+\gamma\\ \leq\mu_{\omega}(Z_{\bar{M}}\cap Z_{\omega})=\int_{Z_{\omega}}1_{Z_{\hat{M}}}(x)d\mu_{\omega}=LF_{Z_{\hat{M}}}(x),
\end{eqnarray}
for all macro-states $M$ with $M\neq \hat{M}$. Hence $\hat{M}$ fulfills the definition of a
$\gamma$-$\varepsilon$-equilibrium.

\subsection{Proof of the Ergodicity-Corollary and the Epsilon-Ergodicity-Corollary}

\noindent \emph{Proof of the Ergodicity-Corollary}:\\
From the definition of an ergodic system (equation \ref{ergodic}), it immediately follows that if an ergodic system has a macro-region $Z_{\hat{M}}$ that is $\beta$-dominant, $\hat{M}$ is an $\alpha$-$\varepsilon$-equilibrium for $\alpha=\beta$. Similarly, from the definition of an ergodic system (equation \ref{ergodic}), it follows that if an ergodic system has a macro-region $Z_{\hat{M}}$ that is $\delta$-prevalent, $\hat{M}$ is a $\gamma$-$\varepsilon$-equilibrium for $\gamma=\delta$.\\

\noindent \emph{Proof of the Epsilon-Ergodicity-Corollary}:\\
An epsilon-ergodic system  $(Z,\Sigma_{Z},\mu_{Z},T_{t})$ is ergodic on a set $\hat{Z}$ with $\mu_{Z}(\hat{Z})=1-\varepsilon$ for a very small $\varepsilon\geq 0$. Hence equation (\ref{ergodic}) implies that if an epsilon-ergodic system has a macro-region $Z_{\hat{M}}$ that is $\beta$-dominant for $\beta-\varepsilon>\frac{1}{2}$, $LF_{Z_{\hat{M}}}(x)\geq \beta-\varepsilon$ for almost all $x\in \hat{Z}$. Consequently, $Z_{\hat{M}}$ is a $\alpha$-$\varepsilon$-equilibrium for $\alpha=\beta-\varepsilon$. Similarly, equation (\ref{ergodic}) implies that if an epsilon-ergodic system has a macro-region $Z_{\hat{M}}$ that is $\delta$-prevalent for $\delta-\varepsilon>0$, $LF_{Z_{\hat{M}}}(x)\geq LF_{Z_{M}}(x)+\delta-\varepsilon$ for all macro-states $M\neq \hat{M}$ and almost all $x\in \hat{Z}$. Hence $Z_{\hat{M}}$ is a $\gamma$-$\varepsilon$-equilibrium for $\gamma = \delta-\varepsilon$.

\section*{REFERENCES}

\noindent Ainsworth, P.M. 2012. ``Entropy in Statistical Mechanics''. \emph{Philosophy of Science} 79: 542-560.\\

\noindent Albert, D. 2000. \emph{Time and Chance}. Cambridge/MA and London: Harvard University Press.\\

\noindent Boltzmann, L. 1877. ``\"{U}ber die Beziehung zwischen dem zweiten Hauptsatze der mechanischen W\"{a}rmetheorie und der Wahrscheinlichkeitsrechnung resp.\ den S\"{a}tzen \"{u}ber das W\"{a}rmegleichgewicht''. \emph{Wiener Berichte} 76: 373-435.\\

\noindent Brown, H. and Uffink, J. 2001. ``The Origins of Time-Asymmetry in Thermodynamics: The Minus First Law''. \emph{Studies in History and Philosophy of Modern Physics} 32: 525-538.\\

\noindent Bricmont, J. 2001. Bayes, Boltzmann and Bohm: Probabilities in Physics. In J. Bricmont, D. D\"{u}rr, M.C. Galvotti, G. Ghirardi, F. Petruccione and N. Zanghi (eds.), \emph{Chance in Physics: Foundations and Perspectives}, 3-21. Berlin: Springer.\\

\noindent Callender, C. 2001. ``Taking Thermodynamics Too Seriously''. \emph{Studies in History and Philosophy of Modern Physics} 32: 539-553.\\

\noindent Dizadji-Bahmani, F., Frigg, R. and Hartmann, S. 2010. ``Who Is Afraid of Nagelian Reduction?''. \emph{Erkenntnis} 73: 393-412.\\

\noindent Ehrenfest, P. and Ehrenfest, T. 1959. \emph{The Conceptual Foundations of the Statistical Approach in Mechanics}. Ithaca, New York: Cornell University Press.\\

\noindent Frigg, R. 2008. ``A Field Guide to Recent Work on the Foundations of Statistical Mechanics.'' In D. Rickles (ed.), \emph{The Ashgate Companion to Contemporary Philosophy of Physics}, 99-196. London: Ashgate.\\

\noindent Frigg, R. 2010a. ``Why Typicality Does Not Explain the Approach to Equilibrium''. In M. Suárez (ed.), \emph{Probabilities, Causes and Propensities in Physics}, 77-93. Dordrecht: Springer. \\

\noindent Frigg, R. 2010b. ``Probability in Boltzmannian Statistical Mechanics''. In G. Ernst and A. Hüttemann (eds.), \emph{Time, Chance and Reduction. Philosophical Aspects of Statistical Mechanics}. Cambridge: Cambridge University Press.\\

\noindent Frigg, R. and Werndl, C. 2012a. ``Demystifying Typicality.'' \emph{Philosophy of Science} 79:917-929.\\

\noindent Frigg, R. and Werndl, C. 2012b. ``A New Approach to the Approach to Equilibrium''.  In Y. Ben-Menahem and M. Hemmo (eds.), \emph{Probability in Physics}, 99-114. The Frontiers Collection. Berlin: Springer.\\

\noindent Frigg, R. and Werndl, C. 2011a. ``Explaining Thermodynamic-Like Behaviour in Terms of Epsilon-Ergodicity''. \emph{Philosophy of Science} 78: 628-652.\\

\noindent Frigg, R. and Werndl, C 2011b. ``Entropy -- A Guide for the Perplexed''. In: C. Beisbart and S. Hartmann (eds.), \emph{Probabilities in Physics}, 115-142. Oxford: Oxford University Press.\\

\noindent Goldstein, S. 2001. ``Boltzmann's Approach to Statistical Mechanics''. In J. Bricmont, D. D\"{u}rr, M. Galavotti, G. Ghirardi, F. Pettrucione and N. Zanghi (eds.), \emph{Chance in Physics: Foundations and Perspectives}, 39-54. Berlin
and New York: Springer.\\

\noindent Goldstein, S. and Lebowitz, J.L. 2004. ``On the (Boltzmann) Entropy of Nonequilibrium
Systems''. \emph{Physica D} 193:53-66.\\

\noindent Gupta, M.C. 2003. \emph{Statistical Thermodynamics}. New Delhi: New Age International Publishing.\\

\noindent Lavis, D. 2005. ``Boltzmann and Gibbs: An Attempted Reconciliation''. \emph{Studies in History and Philosophy of Modern Physics} 36: 245-273.\\

\noindent Lavis, D. 2008. ``Boltzmann, Gibbs and the Concept of Equilibrium''. \emph{Philosophy of Science} 75:682-696.\\

\noindent Lebowitz, J.L. 1993a. ``Boltzmanns Entropy and Time's Arrow''. \emph{Physics Today} (September), 32-38.\\

\noindent Lebowitz, J.L. 1993b. ``Macroscopic Laws, Microscopic Dynamics, Time's Arrow and Boltzmann's Entropy". \emph{Physica A} 194, 1-27.\\

\noindent MacDonald, D.K.C (1962). \emph{Noise and Fluctuations}. An Introduction. Wiley. \\

\noindent Penrose, R. 1989. \emph{The Emperor's New Mind}. Oxford: Oxford University Press.\\

\noindent Petersen, K. 1983. \emph{Ergodic theory}. Cambridge: Cambridge University Press.\\

\noindent Reiss, H. 1996. \emph{Methods of Thermodynamics}. Mineaola/NY: Dover.\\

\noindent Uffink, J. 2001. ``Bluff Your Way in the Second Law of Thermodynamics''. \emph{Studies in History and Philosophy of Modern Physics} 32: 305-394.\\

\noindent Uffink, J. 2007. Compendium of the Foundations of Classical Statistical
Physics. In J. Butterfield and J. Earman (eds.), \emph{Philosophy of Physics}, 923-1047. Amsterdam: North Holland.\\

\noindent Werndl, C. 2013. ``Justifying Typicality Measures in Boltzmannian Statistical Mechanics''. \emph{Studies in History and Philosophy of Modern Physics} 44: 470-479.\\

\noindent Werndl, C. 2009. `'Are Deterministic Descriptions and Indeterministic Descriptions Observationally Equivalent?''. \emph{Studies in History and Philosophy of Modern Physics} 40, 232-242.\\

\end{document}